\documentclass{elsart}

\usepackage{epsfig}

\usepackage{amssymb}

\begin{document}

\begin{flushright}
FERMILAB-Conf-01/240-E
\end{flushright}

\begin{frontmatter}

% Title, authors and addresses

% use the thanksref command within \title, \author or \address for footnotes;
% use the corauthref command within \author for corresponding author footnotes;
% use the ead command for the email address,
% and the form \ead[url] for the home page:
% \title{Title\thanksref{label1}}
% \thanks[label1]{}
% \author{Name\corauthref{cor1}\thanksref{label2}}
% \ead{email address}
% \ead[url]{home page}
% \thanks[label2]{}
% \corauth[cor1]{}
% \address{Address\thanksref{label3}}
% \thanks[label3]{}

\title{Muon Cooling R\&D}

% use optional labels to link authors explicitly to addresses:
% \author[label1,label2]{}
% \address[label1]{}
% \address[label2]{}

\author{Steve Geer\thanksref{label1}}
\thanks[label1]{Invited talk at the NUFACT01 Workshop, Tsukuba, Japan, 
24-30 May 2001.}

\address{Fermi National Accelerator Laboratory, P.O. Box 500, Batavia, IL 60510, USA}

\begin{abstract}
International efforts are under way to design and test a muon 
ionization cooling channel. The present R\&D program is described, and 
future plans outlined.
\end{abstract}

\end{frontmatter}

% main text

\vspace{-0.5cm}

\section{Introduction}
\label{Intro}

Recently there has been widespread interest in developing a 
very intense muon source capable of producing a millimole of muons per 
year. If the muons are then accelerated to high 
energies, they could be used in a Neutrino Factory~\cite{sgprd}, 
and perhaps eventually a Muon Collider~\cite{mucollider}. 

The muons are to be produced using an intense proton source to make 
low energy charged pions, which are confined within  
a large acceptance decay channel. The daughter muons 
produced from $\pi^\pm$ decays will occupy a large phase-space volume. 
Before the muons can be accelerated   
the transverse phase-space they occupy must be reduced 
so that the muon beam fits within the acceptance of an accelerator. 
This means we must ``cool'' the 
transverse phase--space by at least a factor of a few in each 
transverse plane. This must be done fast, before the muons decay. 
Stochastic-- and electron--cooling are too slow. 
It is proposed to use a new cooling technique, namely 
``ionization cooling''~\cite{cool}. 

In an ionization cooling channel the muons 
pass through an absorber in which they lose transverse-- and 
longitudinal--momentum by {\it dE/dx} losses. The longitudinal momentum 
is then replaced using an RF cavity, and the process is 
repeated many times, removing the transverse muon momenta. This  
cooling process will compete with transverse heating due to 
Coulomb scattering. To minimize the effects of scattering we chose 
low--Z absorbers placed in the cooling channel 
lattice at positions of low--$\beta_\perp$ so that the typical radial 
focusing angle is large. If the focusing angle is much larger 
than the average scattering angle then scattering will not have much 
impact on the cooling process.

\begin{figure}%1
%\vspace{-2.0cm}
\centering
\epsfxsize300pt
%\epsffile{sfofo1.ps}
\epsffile{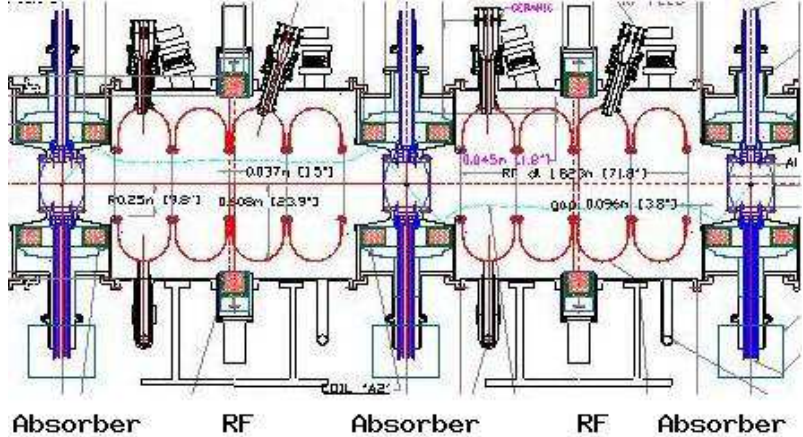}
%\vspace{-6.5cm}
\caption{SFOFO cooling channel design. A 5.5~m long section 
is shown, consisting of two 
200 MHz four-cell cavities interleaved with three liquid 
hydrogen absorbers.}
\label{fig:sfofo}
\end{figure}
%
%
%\begin{figure}%1
%\epsfxsize180pt
%\centering
%\epsffile{dflip.ps}
%\caption{ }
%\label{fig6}
%\end{figure}

\section{Cooling Channel Design in the US}

In the last 18 months there have been two Neutrino Factory ``Feasibility'' 
Studies~\cite{studies} in the US which have involved end-to-end design 
studies, 
together with detailed simulations for each piece of the Neutrino Factory 
complex. The studies have used two simulation tools developed by the 
Neutrino Factory and Muon Collider Collaboration: (i) A specially developed 
tracking code ICOOL, and (ii) A GEANT based program with accelerator 
components (e.g. RF cavities) implemented. Out of these design and 
simulation studies, two promising cooling channel designs have emerged: 
\begin{description}
\item{(i)} The ``SFOFO'' lattice in which the absorbers are 
located at low--$\beta_\perp$ locations 
within high-field solenoids. The field rapidly decreases from a 
maximum to zero at the absorber center, and then increases to a maximum 
again with the axial field direction reversed. 
Figure~\ref{fig:sfofo} shows the design for a 5.5~m long section of the 
$\sim 100$~m long cooling channel. The section shown has 30~cm long 
absorbers with a radius of 15~cm, within a 3.5~T axial field. 
Towards the end of the cooling channel the maximum field is 
higher (5~T) and the lattice period shorter (3.3~m).  
The RF  
cavities operate at 200~MHz and provide a peak gradient of 17~MV/m. 
Detailed simulations predict that the SFOFO channel increases  
the number of muons within the accelerator acceptance  
by a factor of 3-5 (depending 
on whether a large- or very-large acceptance accelerator is used).
\item{(ii)} The ``DFLIP'' lattice in which the solenoid field remains constant
over large sections of the channel, reversing direction only twice. 
In the early part of the channel the muons lose mechanical angular momentum 
until they are propagating parallel to the axis. After the first field 
flip the muons have, once again, mechanical angular momentum, and 
hence move along helical trajectories with Lamour centers along the 
solenoid axis. Further cooling removes the mechanical angular momentum, 
shrinking the beam size in the transverse directions. 
The field in the early part of the channel is 3~T, increasing to 7~T for
the last part. Detailed simulations 
show the performances of the DFLIP and SFOFO channels are comparable.
\end{description}
Earlier less detailed studies~\cite{mucollider} 
have shown that a much larger cooling factor will be required for a 
muon collider. This will require an extended cooling channel, using 
higher frequency (e.g. 805 MHz) cavities and higher field solenoids.

\section{MUCOOL R\&D}

The mission of the MUCOOL collaboration is to design, 
prototype, and bench--test all cooling channel components, and eventually 
beam--test a cooling section. The main component issues are 
(i) can sufficiently high gradient RF cavities be built and operated 
in the appropriate magnetic field and radiation environment, 
(ii) can liquid hydrogen absorbers with thin enough windows be built 
so that the $dE/dx$ heating can be safely removed, and (iii) can the 
lattice solenoids be built to tolerance and be affordable? 
The MUCOOL collaboration 
has embarked on a design--, prototyping--, and testing--program for 
all these components. This is expected to proceed over the next 
3~years.
\begin{figure}%1
%\vspace{-2.0cm}
\centering
\epsfxsize250pt
%\epsffile{solencavity1.ps}
\epsffile{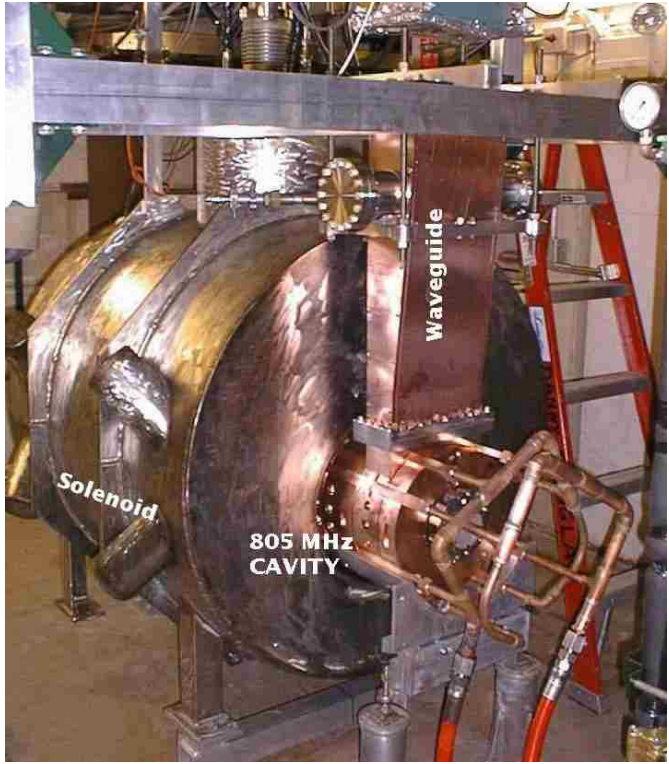}
%\vspace{-6.5cm}
\caption{High power 805 MHz cavity test in the MUCOOL Lab G test 
area, Fermilab.}
\label{fig:labg}
\end{figure}

\subsection{805 MHz RF Tests}

Early design work for a Muon Collider showed that 
the cooling channel requires 805 MHz cavities operating in a 5T 
solenoid, and providing a peak gradient  
of $\sim30$~MV/m. This deep potential well is needed to keep the 
muons bunched as they propagate down the channel. 
This requirement led to two cavity concepts: 
(a) an open cell design, and (b) a design in which the 
penetrating nature of the muons is exploited by closing the RF aperture 
with a thin conducting Be window (at fixed peak power this doubles 
the gradient on axis).
 
The MUCOOL collaboration has pursued an aggressive 805 MHz cavity 
development program, which is now advanced. The main results to date are:
(i) A 12 MW high power test facility has been built and operated at 
Fermilab (Lab G). The Lab G facility enables 
805 MHz cavities to be tested  within a 5T solenoid. 
(ii) An open cell cavity suitable for a muon cooling channel has been 
designed, an aluminum model built and measured, and a prototype copper 
cavity built, tuned, and successfully tested at full power in the Lab G 
facility. 
(iii) A Be foil cavity has been designed at LBNL, a low power test cavity 
built and measured, and foil deflection studies made to ensure the cavity 
does not detune when the foil is subject to RF heating. A high power copper 
cavity with Be-foil windows is under construction at LBNL and the 
University of Mississippi, and will be tested at Lab G when ready.
 
\subsection{200 MHz Cavity Development}

The cooling channel designs developed for the US Neutrino Factory studies 
require 200 MHz RF cavities providing a gradient on axis of $\sim 17$~MV/m. 
Preliminary cavity designs have been made. There are two concepts, 
both of which close the cavity aperture. The options are to use (a) a thin 
Be foil, exploiting the work done for the 805 MHz cavity, or 
(b) use a grid of hollow conducting tubes. Preliminary mechanical tests 
for both the grid and foil concepts are planned, and should proceed 
during the next few months. A 200 MHz prototype cavity will then be 
constructed, and should be ready for high power tests in about 2 years.
\begin{table}[]
\caption{LH$_2$ absorber parameters in Neutrino Factory design study II.}
\begin{center}
%\vspace{0.6 cm}
\begin{tabular}{l|cccccc}
\hline \hline
 &Length&Radius&Number&Heat (kW)&Window Thick&Max. Pres-\\
 Absorbers&(cm)  &(cm)  &Needed&Deposited&-ness ($\mu$m)&sure (atm)\\
\hline
Early&35&18&16&$\sim 0.3$&360&1.2\\
Late &21&11&36&$\sim 0.1$&220&1.2\\
\hline\hline
\end{tabular}
\label{tab:1}
\end{center}
\end{table}
\subsection{Absorber Development}

The cooling channel liquid hydrogen absorbers must have 
very thin windows to minimize multiple scattering, and must tolerate 
heating of O(100~W) from the ionization energy deposited by the 
traversing muons. Absorber parameters for the Neutrino Factory 
study II cooling channel design are listed in Table~\ref{tab:1}.

To adequately remove the heat from the absorbers requires transverse 
mixing of the liquid hydrogen. There are two design concepts that are 
being pursued: (i) Forced flow design. The LH$_2$ is injected into the 
absorber volume through nozzles, and cooled using an external 
loop and heat exchanger. (ii) Convection design. Convection 
is driven by a heater
at the bottom of the absorber volume, and heat removed by a heat exchanger
on the outer surface of the absorber. A forced flow absorber 
prototype has been designed at the Illinois Institute of Technology (IIT) 
and is under construction. A convection prototype has been 
designed by IIT, KEK, and the University of Osaka, and is under construction 
in Japan. Both absorbers will be tested at Fermilab when complete.

A first prototype 15~cm radius aluminum absorber window has been made at the 
University of Mississippi on a CNC milling machine and lathe. The 
window has a central thickness of 130~$\mu$m. The window thickness 
and profile were measured at FNAL and found 
to be within 5\% of the nominal envelope. This 
verifies the manufacturing procedure. The window has been tested 
under pressure in a setup at Northern Illinois University 
in which it was mounted on a backplate 
and water injected between window and plate. Strain gauge and 
photogrammetric 
measurements were made as a function of pressure, and the results compared 
with FEA predictions. Onset of inelastic deformation was predicted at 29~psig, 
a pinhole leak appeared at 31~psig, and rupture occurred at 44~psig. 
The windows required for a cooling channel absorber can be about 
twice as thick as the first prototype window. The results to date are 
therefore encouraging. Further window studies and tests are proceeding.
\begin{figure}%1
%\vspace{-2.0cm}
\centering
\epsfxsize350pt
%\epsffile{window_test.ps}
\epsffile{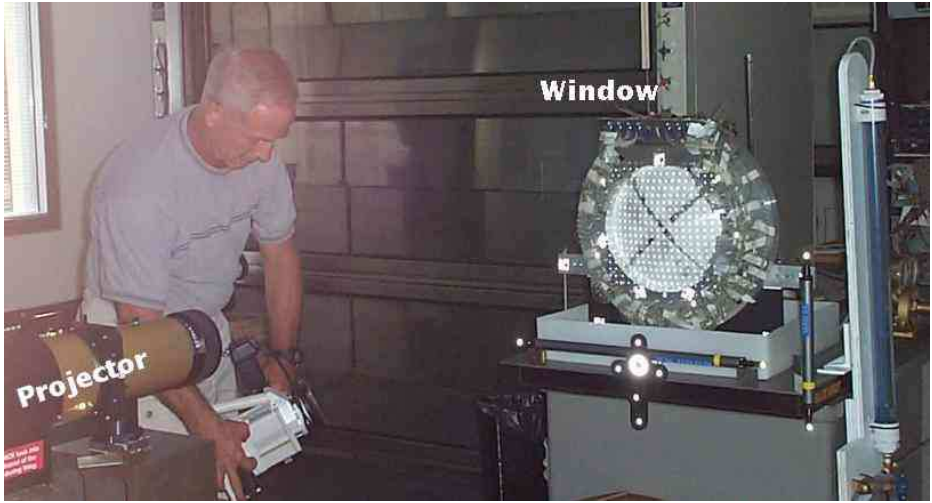}
\vspace{0.5mm}
\caption{Absorber window test, showing an array of dots 
projected onto the window for photogrammetric measurements of its 
shape as it deforms under pressure.}
\label{fig:absorber}
\end{figure}

\section{European R\&D Program}

The CERN cooling channel design is similar in concept to the US design, 
but is based on 44~MHz and 88~MHz cavities rather than 200~MHz cavities. 
To minimize the radii of the solenoids used to confine the muons 
within the channel, the cavities have been designed to wrap around 
the solenoids. A full engineering design of this concept will be 
required to understand its feasibility. 
The initial transverse cooling is performed using 44~MHz cavities with 
four 1~m long RF cells between each 24~cm long LH$_2$ absorber. 
The beam is then accelerated from 200~MeV to 300~MeV, and the cooling 
is continued using 88~MHz cavities with eight 0.5~m long cells 
between each 40~cm long LH$_2$ absorber. The channel parameters are 
summarized in Table~\ref{tab:2}. Simulations of the channel performance 
with detailed field-maps have not yet been made. However, simulations 
using simpler field maps yield promising results: the effect of the 
channel is to increase the number of muons within the acceptance of the 
subsequent accelerating system by a factor of about 20. Whether this 
increased yield is significantly degraded when full simulations are 
performed remains to be seen. In the meantime, a prototype 88~MHz cavity 
is being prepared at CERN for high power tests within the coming year. 
\begin{table}
\caption{\label{tab:2} CERN cooling channel design parameters. 
}
\begin{center}
%\vspace{0.6 cm}
\begin{tabular}{l|cc}
\hline \hline
 &Channel 1&Channel 2\\
\hline
Length& 46~m& 112~m\\
Diameter&60~cm&30~cm\\
Sol. Field&2.0~T&2.6~T\\
RF Freq.&44~MHz&88~MHz\\
RF Gradient&2~ MV/m&4~MV/m\\
Beam Energy&200~MeV&300~MeV\\
\hline\hline
\end{tabular}
\end{center}
\end{table}

\section{Cooling Experiments}

A sequence of muon cooling-related experiments is being planned. 
The first, the MUSCAT experiment~\cite{muscat}, 
is already under way at TRIUMF. 
The second, the MUCOOL Component Test Experiment, is under 
construction at the Fermilab LINAC. The third, an International 
Cooling Experiment~\cite{ice}, is in the planning stage. The fourth, an 
eventual String Test Experiment, will be planned in the future.

\subsection{MUSCAT}

The goal of the MUSCAT experiment at TRIUMF is the precise measurement of 
low energy (130, 150, and 180~MeV/c) 
muon scattering in a variety of materials that might be found 
with a cooling channel. In a second phase, the experiment will also 
measure straggling. Scattering measurements for Li, Be, 
C, Al, CH$_2$, and Fe have already been made. Preliminary results 
seem to be in good agreement with expectations. Further analysis 
is in progress. Measurements with LH$_2$ are expected in the 
future.

\subsection{The MUCOOL Component Test Experiment}

A MUCOOL test area located at the end of the Fermilab 400~MeV LINAC 
was proposed in the Fall of 2000, and is currently under construction. 
The project is being pursued in two phases. In Phase 1 a LH$_2$ absorber 
test facility is being built, which will enable the first prototype 
absorbers to be filled. In Phase 2 the test area will be enlarged, 
a LINAC beam will be brought to the absorber area, and the 5T solenoid 
will be moved from Lab G so that the absorber can be tested in a magnet 
whilst exposed to a proton beam. The beam intensity and spot size 
will be designed to mimic the total ionization energy deposition 
and profile that corresponds to the passage of $10^{12} - 10^{13}$ muons 
propagating through a cooling channel. In addition, 200~MHz RF power 
will be piped to the test area from a nearby test-stand, enabling 
high-power tests to be made of a prototype 200 MHz cooling channel 
cavity exposed to the proton beam.

\subsection{The International Cooling Experiment}

A Europe-Japan-US International Cooling Experiment is currently being 
planned. The goals are to (i) place a cooling channel section (capable of 
achieving the performance required for a Neutrino Factory) in a muon 
beam, and (ii) demonstrate our ability to precisely simulate the passage 
of muons confined within a periodic lattice as they pass through 
LH$_2$ absorbers and high-gradient RF cavities. In the envisioned 
experiment muons are tracked one at a time at the input and output of the 
cooling section, and the precise response of the muons to the cooling 
section is determined. The main challenge to the design of this type 
of experiment arises from the prolific x-ray environment created by the 
RF cavities. This is currently under study at Lab~G and elsewhere. 
If it is found that single particle detectors can function in this 
hostile environment, then we can anticipate a proposal being submitted 
sometime in 2002.

\section{Summary}

Detailed simulations show that a muon ionization cooling channel for 
a Neutrino Factory is feasible provided RF cavities and LH$_2$ absorbers 
can be built to meet agressive performance specifications. 
A component R\&D program is underway to prototype and test 
cooling channel components. An International Cooling Experiment proposal 
is being prepared. If there is adequate support for this R\&D, in a few 
years we should be in a position to chose the best cooling channel design, 
based on components with verified performance.

\end{document}